\documentclass[preprint,12pt]{elsarticle}

\usepackage{graphics}
\usepackage{graphicx}
\usepackage{amssymb}
\journal{Nuclear Physics A}

\newcommand{\heag}{$^3$He($\alpha,\gamma$)$^7$Be }

\begin{document}

\begin{frontmatter}



\title{Activation measurement of the \heag reaction cross section at high energies}


\author[ATOMKI]{C. Bordeanu\corref{onleave}}
\author[ATOMKI]{Gy. Gy\"urky\corref{cor}}
\ead{gyurky@atomki.hu} 
\author[ATOMKI]{Z.~Hal\'asz}
\author[ATOMKI]{T.~Sz\"ucs}
\author[ATOMKI]{G.G.~Kiss}
\author[ATOMKI]{Z.~Elekes}
\author[ATOMKI]{J.~Farkas}
\author[ATOMKI]{Zs.~F\"ul\"op}
\author[ATOMKI]{E.~Somorjai}
\address[ATOMKI]{Institute for Nuclear Research (Atomki), H-4001 Debrecen, POB.51., Hungary} 
%\address[Bucharest]{Horia Hulubei Institute of Physics and Nuclear Engineering (IFIN-HH), 407 Atomistilor, Magurele-Bucharest 077125, Romania.}
\cortext[onleave]{On leave from Horia Hulubei Institute of Physics and Nuclear Engineering (IFIN-HH), Str. Reactorului no. 30, P.O.BOX MG-6, Bucharest - Magurele, Romania} 
\cortext[cor]{corresponding author} 

\begin{abstract}
The astrophysically important \heag reaction was studied at high energies where the available experimental data are in contradiction. A thin window $^3$He gas cell was used and the cross section was measured with the activation method. The obtained cross sections at energies between E$_{c.m.}$\,=\,1.5 and 2.5\,MeV are compared with the available data and theoretical calculations. The present results support the validity of the high energy cross section energy dependence observed by recent experiments. 

\end{abstract}

\begin{keyword}
nuclear astrophysics \sep pp-chain \sep cross section measurement \sep $^3$He($\alpha,\gamma$)$^7$Be reaction

\end{keyword}

\end{frontmatter}

% \linenumbers

\section{Introduction}
\label{sec:intro}

The \heag reaction is an excellent example of those reactions which preserve their importance over several decades despite many experimental and theoretical studies devoted to them. The crucial role played by the \heag reaction in solar hydrogen burning was realized more than 50 years ago, when H.D. Holmgren and R.L. Johnston measured its cross section \cite{hol59} and found a value being about a factor of 100 higher than estimated theoretically \cite{sal52}. This result indicated that the 2nd and 3rd pp-chains \cite{ili07} can compete favorably with the first chain at the temperature of the solar core.

Several experimental studies followed this pioneering work in the next decades aiming at the determination of the \heag cross section at energies as low as possible in order to approach the astrophysically relevant energy range \cite{par63,nag69,osb82,kra82,rob83,vol83,ale84,hil88}. This energy range, the so called Gamow window, lies between about 15 and 30\,keV at 15 MK temperature relevant for the solar core. The cross section at these energies is too small to be measured directly, theory-based extrapolations are therefore necessary to obtain the cross section at solar energies and hence the reaction rate. 

\begin{figure}
\centering
\resizebox{0.8\textwidth}{!}{\includegraphics{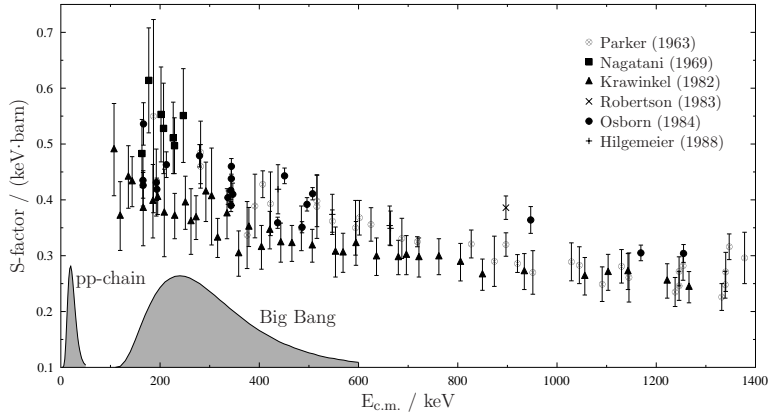}}
\caption{\label{fig:old_data} Experimental data for the \heag S-factor available at the end of the 20$^{th}$ century. The data are from \cite{par63,nag69,osb82,kra82,rob83,hil88}. The Gamow-windows relevant for the solar hydrogen burning via the pp-chain and for the big-bang nucleosynthesis are also shown at the bottom of the figure.}
\end{figure}

Figure\,\ref{fig:old_data} shows the summary of the experimental results available at the time when the first review paper on solar fusion reactions by E.G. Adelberger \textit{et al.} \cite{ade98} was published. In order to avoid the strong energy dependence of the cross section caused by the Coulomb barrier penetration, the results are shown in the form of the astrophysical S-factor\footnote{For the definition of the S-factor see e.g. Ref.\,\cite{ili07} page 172.}. Based on these data, a recommended value of the zero energy extrapolated S-factor of S(0)\,=\,0.53\,$\pm$\,0.05\,keV\,b is quoted in Ref.\,\cite{ade98}. The relative high uncertainty of this value is on one hand due to the typically low precision of the individual measurement at low energies and the large scatter of the data points (see Fig.\,\ref{fig:old_data}). On the other hand, the uncertainty is also increased by the fact that measurements carried out with two different experimental techniques (activation and in-beam methods, see below) resulted in different S(0) extrapolated values \cite{ade98}.

At around the turn of the century the detection of solar neutrinos entered a precision era. The solar neutrino problem has been solved by the discovery of neutrino oscillation \cite{fuk01,ahm01} and the flux of $^7$Be and $^8$B solar neutrinos can now be measured with a precision of about 3-4\,\% by various neutrino detectors \cite{hos06,aha08,bel11,abe11}. The measured neutrino fluxes can therefore be used to probe the solar core and test solar models if the underlying nuclear physics input, i.e. the cross section of pp-chain reactions is known with sufficient precision \cite{bah06}. Since the \heag reaction is the starting point of the 2nd and 3rd chains from where the $^7$Be and $^8$B solar neutrinos originate, the uncertainty of its cross section must be reduced.

The need for a more precise \heag cross section arose also from another field of nuclear astrophysics, namely the big-bang nucleosynthesis. Owing to the high precision measurement of some cosmological parameters, like the baryon-to-photon ratio of the universe carried out by the WMAP mission \cite{wmap}, primordial abundances of some light nuclei can be compared to observation with high accuracy. While good agreement is found between model predictions and observations for most of the nuclei produced in the big-bang, $^7$Li is significantly overproduced in the models. This so called $^7$Li problem is one of the biggest unresolved problems of big-bang theory \cite{ioc09}. At the observed values of the baryon-to-photon ratio of the universe the \heag reaction is responsible mainly for the production of $^7$Li. The $^7$Be decays by electron capture to $^7$Li after the primordial nucleosynthesis ceases\footnote{At lower values of the baryon-to-photon ratio $^7$Li would be mainly produced directly by the $^3$H($\alpha,\gamma$)$^7$Li reaction.}. The precise knowledge of the \heag reaction rate is crucial in order to exclude a solution of the $^7$Li problem based on this reaction rate.

\begin{figure}
\centering
\resizebox{0.8\textwidth}{!}{\includegraphics{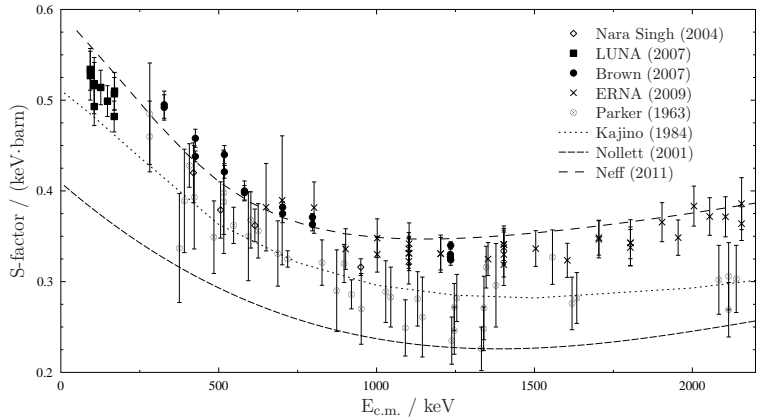}}
\caption{\label{fig:new_data} ``Modern'' experimental data for \heag with total uncertainties. The data are from \cite{nar04,bem06,gyu07,con07,bro07,cos08,dil08}. The results of three microscopic calculations by T. Kajino and A. Arima \cite{kaj84}, K.M. Nollett \cite{nol01} and T. Neff \cite{nef11} are also shown. The discrepancy between different experimental datasets and different calculations is clearly seen and gives the motivation of the present work.}
\end{figure}

Answering the urgent call from these two fields of nuclear astrophysics, several new experiments studied the \heag reaction in the last 10 years \cite{nar04,bem06,gyu07,con07,bro07,cos08,dil08}. The results of these ``modern'' \heag experiments are shown in Fig.\,\ref{fig:new_data}. Not surprisingly these experiments tried to concentrate on the low energy range in order to be as close to astrophysical energies as possible. Only the ERNA collaboration using a novel approach of a recoil separator measured the cross section in a wider energy range up to E$_{c.m.}$\,=\,3.1\,MeV \cite{dil08}. Above about 1.3\,MeV the only dataset previously available was that of P.~Parker and R.~Kavanagh \cite{par63}. Fig.\,\ref{fig:new_data} also includes these data. As one can see, there is a clear disagreement between the two high energy datasets. The ERNA data are significantly higher than the P.~Parker and R.~Kavanagh data and the energy dependence of the S-factor is also different.

This contradiction has a strong astrophysical consequence since the cross section must be determined at solar energies, well below the lowest available measurement and the extrapolations are based on theoretical calculations. The goodness of these theoretical models and thus the reliability of the extrapolation is assessed based on the comparison with experimental data at high energies where data are available. Moreover, the calculated S-factors are often fitted  to available experimental data by a single rescaling parameter to obtain the extrapolated S-factor, even though such a rescaling does not have a strong physical justification for microscopic models \cite{ade11}. As an example, Fig.\,\ref{fig:new_data} shows three theoretical cross section curves which are used to obtain cross sections at the Gamow window. The result of the microscopic models of T. Kajino and A. Arima \cite{kaj84} and K.M. Nollett \cite{nol01} were published when only the ``old'' experimental data were available. In lack of other data, their aim was to describe the data of P.~Parker and R.~Kavanagh at high energies. After the ERNA data became available, T. Neff carried out microscopic model calculations and was able to describe relatively well the ``modern'' experimental data \cite{nef11}. The zero energy extrapolated S-factors from the models differ by as much as 50\,\% (without any rescaling to experimental data). Several other theoretical models are available for the \heag reaction (see e.g. Refs.\,\cite{tom63,kim81,buc85,buc88,lan86,mer86,moh93,dub95,cso00,des04,moh09}). The application of different theoretical models introduces a non-negligible uncertainty in the extrapolated S-factor. The new review paper on solar fusion reactions by E.G.~Adelberger \textit{et al.} \cite{ade11} assigns a theoretical uncertainty of 4\,\% to this value. In order to reduce this uncertainty, the high energy behavior of the \heag cross section must also be understood.

\section{The aim of the present work and the chosen method}
\label{sec:aim}

Presently there is no explanation to the difference between the two datasets at high energies. It is therefore necessary to carry out new experiments in the problematic energy range in order to provide more data for a better understanding of the cross section behavior. The aim of the present work is thus to measure the cross section in the energy range between E$_{c.m.}$\,=\,1.5\,MeV and 2.5\,MeV.

There are two different ways to determine the \heag cross section which were both used several times\footnote{A third method is the direct counting of the created $^7$Be isotopes in a recoil separator. This technique was used up to now solely by the ERNA collaboration. A similar experiment is in preparation at the DRAGON facility at TRIUMF, Vancouver, Canada \cite{sju13}.}. The in-beam technique is based on the detection of prompt $\gamma$-radiation emitted when the fusion of the two Helium isotopes takes place in the direct capture process. The capture can lead to the ground or first excited states of $^7$Be and with the in-beam method partial cross section leading to these two states can be measured separately. The sum of the partial cross sections provides the total cross section. The other method is the activation which exploits the fact that the produced $^7$Be isotope is radioactive. It decays with a half-life of 53.22\,$\pm$\,0.06\,d \cite{til02} and the electron capture decay is followed by the emission of a single $\gamma$-radiation of 478\,keV with 10.44\,$\pm$\,0.04\,\% relative intensity \cite{til02}. The off-line detection of this $\gamma$-radiation can also be used to determine the cross section. This method provides directly the total cross section but gives no information about the branching ratio between the population of the ground state and first excited state.

The set of ``old'' experiments showed an apparent discrepancy of about 15\,\% between the results obtained with the two methods. The new experiments of the LUNA collaboration \cite{bem06,gyu07,con07,cos08} and Brown \textit{et al.} \cite{bro07} provided data with both methods and perfect agreement between the results were found. Therefore, in the present work only one technique was used. The activation method has some clear advantages over the in-beam technique, especially at higher energies. There is no need to determine the angular distribution of the emitted $\gamma$-rays - as opposed to the in-beam method - and the beam induced background is of less importance. Therefore, the activation method has been chosen for the present work.

Since the \heag reaction takes place between two isotopes of a gaseous element, the application of a gas target is necessary. At high energies, a thin window gas cell is a possible choice because the beam energy loss in the entrance foil is typically low compared to the beam energy and does not introduce a high uncertainty in the determination of the effective energy. A thin window gas cell is technically easier to handle and the determination of the number of target atoms can be done with high precision. The present study was thus carried out using a thin window gas cell. Further details about possible gas target designs can be found in \cite{bor12}.

\section{Experimental details}
\label{sec:exp}

The experimental procedure used to determine the \heag cross section has been described in detail in a separate publication \cite{bor12}. Here only the most important aspects of the experiments are summarized.

The $\alpha$-beam was provided by the MGC-20 cyclotron of ATOMKI \cite{cyclotron}. After many tests, five irradiations were carried out between $\alpha$-energies of 4.0 and 6.3 MeV. The energy of the beam at the used beamline is defined to a precision of 0.3\,\%. The typical $^4$He$^{++}$ beam intensity was between 100 and 250\,pnA. The length of irradiations varied between 14 and 25 hours (see table\,\ref{tab:irrad}). Although the irradiation time was short compared to the half-life of the reaction product, the current integrator counts were recorded in multichannel scaling mode with 1 min time base in order to follow the changes in beam intensity. The current integrator was calibrated using a current generator and a precision current meter. 

The beam entered the target chamber through a beam defining aperture of 4\,mm in diameter. After this aperture the whole chamber, including the gas cell, served as a Faraday cup in order to determine the number of projectiles impinging on the target by charge measurement. A larger diameter aperture was placed behind the beam defining aperture and biased to -300\,V to suppress the secondary electrons emitted from the target and aperture. A liquid nitrogen cooled trap was placed upstream the chamber to reduce the carbon build-up on the gas cell entrance foil.

The $^3$He target gas was confined in a closed gas cell of 3\,cm length. The $^3$He isotopic enrichment was 99.95\,\%. The gas was filled into the cell before the irradiation and kept closed afterwards. The pressure of the gas was continuously measured with a type 722B MKS Baratron. The gas pressure was typically 300\,mbar. The temperature of the gas cell was measured with a thermometer attached to the body of the cell. The back of the gas cell which served as the beam stop was directly water cooled. The typical temperature of the cell was about 25\,$^\circ$C determined by the cooling water temperature.

\begin{table}
\centering
\caption{\label{tab:irrad} Some parameters of the irradiations.}
\begin{tabular}{lcccc}
\hline
Run no.  & E$_\alpha$ & Length of  & Total number & Average target \\
 & [MeV] & irrad. [h] & of projectiles & thickness [atoms/cm$^2$] \\
\hline
\#1 & 4.00 & 24.3 & 7.96$\cdot$10$^{16}$ & 2.16$\cdot$10$^{19}$ \\
\#2 & 4.70 & 24.7 & 9.68$\cdot$10$^{16}$ & 2.33$\cdot$10$^{19}$ \\
\#3 & 5.40 & 13.1 & 6.57$\cdot$10$^{16}$ & 2.17$\cdot$10$^{19}$ \\
\#4 & 5.90 & 16.2 & 6.80$\cdot$10$^{16}$ & 2.34$\cdot$10$^{19}$ \\
\#5 & 6.32 & 22.7 & 5.20$\cdot$10$^{16}$ & 2.35$\cdot$10$^{19}$ \\
\hline
\end{tabular}
\end{table}

The local density reduction of the gas when the beam is passing through, the so called beam heating effect, can alter significantly the effective number of target atoms. In the case of the present work the relatively low beam intensity and the low stopping power at the studied energies makes the beam heating effect rather low. The dissipated power of the beam was typically 1\,mW/mm. M. Marta \textit{et al.} studied extensively the beam heating effect with $\alpha$-beam penetrating a $^3$He target \cite{mar06}. The obtained value for the necessary correction was found to be (0.0091\,$\pm$\,0.0019)\,(mW/mm)$^{-1}$. This translates into a 1\,\% beam heating correction in our case. Previously, J.~G\"orres \textit{et al.} found a value of about a factor of two lower \cite{gor80}. In the analysis of the present experimental data a beam heating correction factor of (1\,$\pm$\,1)\,\% was therefore assumed for all five runs.

The beam enters the gas cell through a 1\,$\mu$m thick pinhole-free Ni entrance foil glued onto a stainless steel frame having an 8\,mm diameter hole. The thickness of the foil was measured with $\alpha$-energy loss and perfect agreement with factory value of 1\,$\mu$m was found \cite{bor12}. Table\,\ref{tab:irrad} lists some parameters of the irradiations. 

According to the reaction kinematics, the created $^7$Be nuclei move forward within a cone of a few degree opening angle. In the studied energy range and with the gas target thickness used, the $^7$Be nuclei have enough energy to reach the end of the cell and be implanted into the beam-stop. A removable Oxygen Free High Conductivity copper disk was used at the beam stop as catcher to collect the created reaction products. The catcher material was chosen based on several test measurements aiming at finding parasitic production of $^7$Be in them. For details, see \cite{bor12}. GEANT4 simulations proved that more than 99.7\,\% of the produced $^7$Be reaches the catcher within a circular spot of 24\,mm in diameter around the beam axis. At the studied energies backscattering loss of $^7$Be from the catcher is well below 1\,\% and it is therefore neglected \cite{bor12}.

\begin{figure}
\centering
\resizebox{0.8\textwidth}{!}{\includegraphics{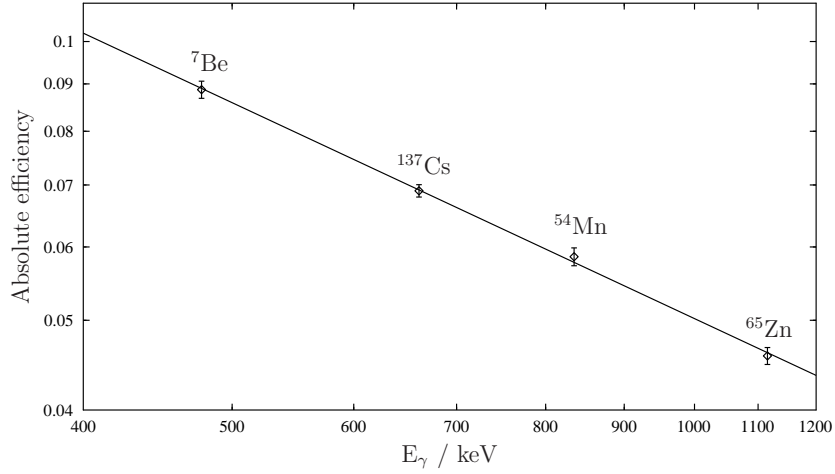}}
\caption{\label{fig:efficiency} Absolute efficiency of the HPGe detector as a function of energy. In order to avoid true coincidence summing, only single line calibration sources were used. The good log-log straight line fit to the measured data strengthens the reliability of the measured efficiency at 478\,keV with the calibrated $^7$Be source.}
\end{figure}

After the irradiations the Cu catcher was removed from the chamber and the collected $^7$Be activity was measured with a 100\,\% relative efficiency HPGe detector equipped with a complete 4$\pi$ low background shielding. In order to maximize the detector efficiency, the catcher were placed in a close geometry in front of the detector. The distance between the catcher and the detector end cap was 1\,cm. Since $^7$Be is a single line $\gamma$-emitter, no true coincidence summing effect is present. In order to avoid coincidence summing also in the case of calibration sources, the absolute efficiency of the $\gamma$-detector was measured with single line sources only. These sources were the following: $^{54}$Mn, $^{137}$Cs, $^{65}$Zn and $^{7}$Be. Fig.\,\ref{fig:efficiency} shows the measured efficiencies with these sources and the fitted curve. In order to take into account the different spatial distribution of the calibration sources and the actual $^{7}$Be source on the catcher, GEANT4 simulations were carried out. The final efficiency at 478\,keV was determined to be (8.87\,$\pm$\,0.44)\,\%.

\begin{figure}
\centering
\resizebox{0.8\textwidth}{!}{\includegraphics{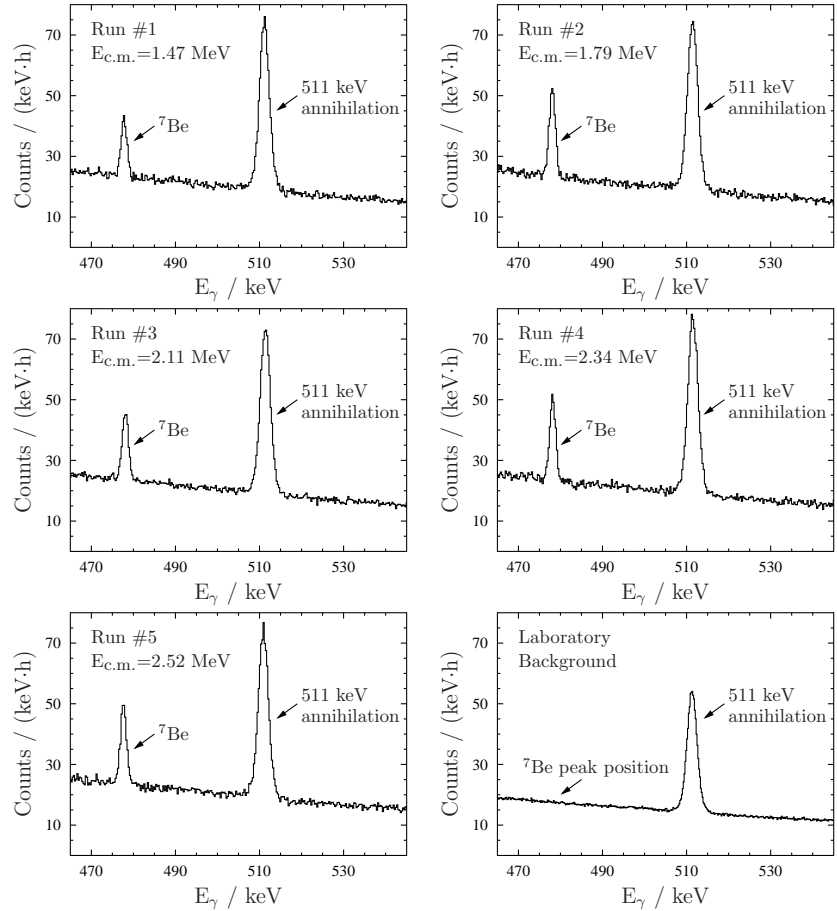}}
%\resizebox{0.45\textwidth}{!}{\rotatebox{270}{\includegraphics{run1_spectrum.eps}}}
%\resizebox{0.45\textwidth}{!}{\rotatebox{270}{\includegraphics{run2_spectrum.eps}}}
%\resizebox{0.45\textwidth}{!}{\rotatebox{270}{\includegraphics{run3_spectrum.eps}}}
%\resizebox{0.45\textwidth}{!}{\rotatebox{270}{\includegraphics{run4_spectrum.eps}}}
%\resizebox{0.45\textwidth}{!}{\rotatebox{270}{\includegraphics{run5_spectrum.eps}}}
%\resizebox{0.45\textwidth}{!}{\rotatebox{270}{\includegraphics{bck_spectrum.eps}}}
\caption{\label{fig:spectra} Relevant part of the measured $\gamma$-spectra in the five runs. In the lower right panel the laboratory background is plotted. The spectra are normalized to counting time.}
\end{figure}

Table\,\ref{tab:counting} lists some parameters of the $\gamma$-countings. Fig.\,\ref{fig:spectra} shows the relevant part of the $\gamma$-spectra of the five runs as well as the laboratory background spectrum. The spectra are normalized to counting time. 
\begin{table}
\centering
\caption{\label{tab:counting} Some parameters of the $\gamma$-countings. The waiting time is the time elapsed between the end of the irradiation and the start of the counting.}
\begin{tabular}{lccc}
\hline
Run no.  & Waiting & Counting  & $^7$Be net\\
 & time [h] & time [h] & area [cts] \\
\hline
\#1 & 1 & 192 & 4559\,$\pm$\,178 \\
\#2 & 106 & 168 & 6422\,$\pm$\,216 \\
\#3 & 8 & 288 & 8630\,$\pm$\,236 \\
\#4 & 81 & 192 & 6681\,$\pm$\,239 \\
\#5 & 63 & 168 & 5661\,$\pm$\,175 \\
\hline
\end{tabular}
\end{table}

The cross section $\sigma$ was calculated using the following formula:
\begin{equation}
		\sigma = \left(T\cdot\phi\cdot\frac{1-e^{\lambda\cdot t_i}}{\lambda}\right)^{-1}\cdot \frac{A\cdot e^{\lambda\cdot t_w}}{(1-e^{-\lambda\cdot t_c})\cdot\epsilon_\gamma\cdot\eta}
\end{equation}
where $T$ [atoms/cm$^2$] is the target thickness, $\phi$ [1/s] is the beam intensity, $\lambda$ is the $^7$Be decay constant, $\epsilon_\gamma$ is the gamma detection efficiency at 478\,keV and $\eta$ is the decay branching to the first excited state in $^7$Li leading to the 478\,keV $\gamma$-radiation. $A$ is the measured 478\,keV net peak area. The time constants $t_i$, $t_w$, $t_c$ are the length of the irradiation, the waiting time between  irradiation and  counting and the length of the counting, respectively.

\section{Results and discussion}
\label{sec:results}

Table\,\ref{tab:results} summarizes the experimental results. The second column shows the effective center-of-mass energies. The effective energies were calculated based on the following steps. The energy loss in the Ni entrance foil obtained from the SRIM code \cite{SRIM} was subtracted from the beam energy. The thickness of the foil was taken to be 1.01\,$\pm$\,0.04\,$\mu$m \cite{bor12}. Then the energy loss in the target gas was considered. Since the energy loss in the gas is rather low (typically around 150\,keV) and the S-factor does not change much within this energy interval, a constant S-factor over the target thickness was assumed and the effective energy was calculated based on the procedure given e.g in chapter 4.8 of \cite{ili07}. Finally, the effective energy was converted into the center-of-mass frame. The uncertainty of the effective energy is obtained by assigning 0.3\,\% uncertainty to the primary beam energy, taking into account the energy straggling and using 3.9\,\% uncertainty for the stopping power values \cite{SRIMerror}. In principle, the energy straggling cannot be treated the same way as the other uncertainties since it represents only the broadening of the beam energy distribution. In order to quote a conservative estimate of the uncertainty of energy, and taking into account its small contribution to the final uncertainty, the value of the energy straggling is, however, included in the calculation of the energy uncertainty.

\begin{table}
\centering
\caption{\label{tab:results} Results of the present work in the form of cross section and S-factor. The statistical and total systematic uncertainties, respectively, are quoted separately in parentheses. See text for further details.}
\begin{tabular}{lccc}
\hline
Run no.  & E$_{c.m.}^{eff.}$ & Cross section  & S-factor\\
 & [keV] & [$\mu$barn] & [eV barn] \\
\hline
\#1 & 1473\,$\pm$\,15 & 2.96(12)(17) & 313(12)(19) \\
\#2 & 1791\,$\pm$\,14 & 3.78(13)(22) & 327(11)(19) \\
\#3 & 2115\,$\pm$\,14 & 4.68(13)(28) & 351(10)(21) \\
\#4 & 2338\,$\pm$\,14 & 5.08(18)(30) & 354(13)(21) \\
\#5 & 2527\,$\pm$\,14 & 6.06(19)(36) & 401(12)(24) \\
\hline
\end{tabular}
\end{table}

The uncertainty of the cross section values is calculated as the quadratic sum of the statistical error of the $^7$Be peak area and several systematic uncertainties. These uncertainties are listed in Table\,\ref{tab:uncert}. In Table\,\ref{tab:results} the statistical and total systematic uncertainties, respectively, are quoted separately in parenthesis. The uncertainty of the S-factor values also includes the effective energy uncertainty, however, since the effective energy is rather well defined, this source of uncertainty is small compared with the others listed in table\,\ref{tab:uncert}.

\begin{table}
\centering
\caption{\label{tab:uncert} List of uncertainties on the cross section (S-factor) determination}
\begin{tabular}{llc}
\hline
Type  & Source of & Value in \\
 & uncertainty & percentage \\
\hline
statistical & $^7$Be peak area & 2.7\,--\,3.9\,\% \\
\hline
						& $\gamma$-detection efficiency incl. source distribution & 4\,\% \\
						& number of target atoms$^*$ & 3\,\% \\
systematic	& current integration & 3\,\% \\
						& beam heating & 1\,\% \\
						& effective energy$^{**}$ & 0.4\,--\,1.1\,\% \\
\hline
\multicolumn{2}{l}{total systematic} & 5.9\,\,\%\\
\hline
\multicolumn{3}{l}{\footnotesize $^*$includes gas pressure, temperature and gas cell length uncertainties} \\
\multicolumn{3}{l}{\footnotesize $^{**}$affecting only the S-factor, not the cross section}
\end{tabular}
\end{table}

\begin{figure}
\centering
\resizebox{0.9\textwidth}{!}{\includegraphics{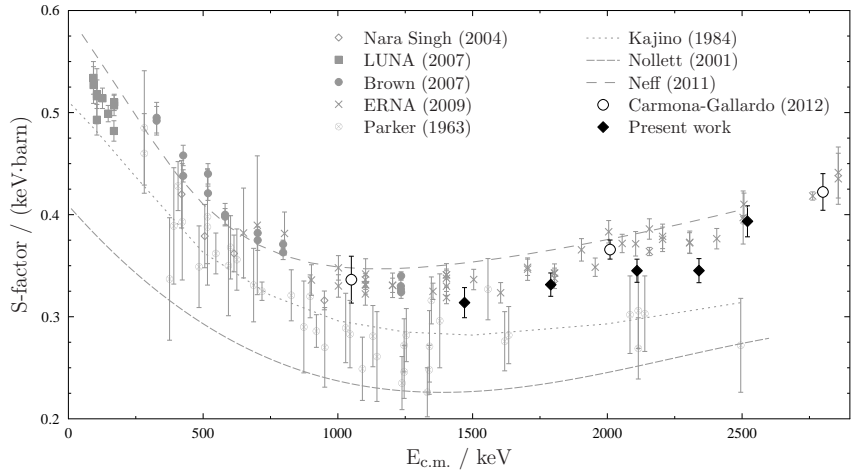}}
\caption{\label{fig:results} Results of the present work compared with ``modern'' data and some theoretical calculations. Only the statistical uncertainties are plotted with the exception of the P.~Parker and R.~Kavanagh data where different sources of uncertainty are not quoted separately. See text for discussion.}
\end{figure}

Figure\,\ref{fig:results} shows the results in the form of S-factor together with the data of the ``modern'' experiments and that of P.~Parker and R.~Kavanagh. Along with the experimental data, the calculated S-factor of the three microscopic models introduced in Fig.\,\ref{fig:new_data} are also plotted without any rescaling.

When the present study was in progress, M.~Carmona-Gallardo \textit{et al.} published their results on \heag in a similar energy range providing cross section values at three energies \cite{car12}. The motivation of that work was similar to ours. The results of this new experiment are also included in Fig.\,\ref{fig:results}. 

The present data support the findings of the two recent high energy datasets by the ERNA collaboration \cite{dil08} and M.~Carmona-Gallardo \textit{et al.} \cite{car12}. The rather low values and flat energy dependence of the P.~Parker and R.~Kavanagh data are not reproduced by the new experiments.

%Taking into account the total uncertainties of the high energy data, however, it is difficult to draw a definite conclusion about the high energy behavior of the \heag cross section. For example, a downscaling of about 8\,\% of the T. Neff calculation would lead to a better agreement with our experimental data as well as with the lower energy data from the LUNA collaboration \cite{bem06,gyu07,con07,cos08} and B.S. Nara Singh \textit{et al.} \cite{nar04}. On the other hand, such a downscaling would worsen the agreement between the calculation and the data from the ERNA collaboration \cite{dil08}, T.A.D. Brown \textit{et al.} \cite{bro07} and M. Carmona-Gallardo, \textit{et al.} \cite{car12}. 

In summary, the results presented in this paper give a significant contribution to the better understanding of the high energy behavior of the \heag cross section. In the review paper of E.G. Adelberger, et al. \cite{ade11} the available \heag data have been critically reviewed, but the S-factor fitting was restricted to energies below 1\,MeV. A global analysis of the experimental data also at higher energies and their comparison with theory would also be needed for a more complete description. Such an analysis is beyond the scope of the present work.

%The precision required for astrophysical applications is, however, not reached yet as the different data scatter within their corresponding uncertainties of several percent. Further high precision data preferably spanning a wide energy range seems therefore still highly needed.

\section*{Acknowledgments}

This work was supported by OTKA grants No. K101328, PD104664 and NN83261. C.B. acknowledges support from the HUMAN MB-08-B Mobility Project NKTH-OTKA-EU FP7 (Marie Curie) 82409. G.G.K. is supported by the Bolyai Grant.


\begin{thebibliography}{00}

\bibitem{hol59} H.D. Holmgren, R.L. Jonhston, Phys. Rev. 113 (1959) 1556.
\bibitem{sal52} E.E. Salpeter, Phys. Rev 88 (1952) 547.
\bibitem{ili07} C. Iliadis, Nuclear Physics of Stars, WILEY-VCH Verlag GmbH and Co. KGaA, Weinheim, Gemany, 2007.
\bibitem{par63} P. Parker, R. Kavanagh, Phys. Rev. 131 (1963) 2578.
\bibitem{nag69} K. Nagatani, M.R. Dwarakanath, D. Ashery, Nucl. Phys. A 128 (1969) 325.
\bibitem{osb82} J. Osborne, et al., Phys. Rev. Lett. 48 (1982) 1664.
\bibitem{kra82} H. Krawinkel, et al., Zeitschrift f\"ur Physik A 304 (1982) 307.
\bibitem{rob83} R.G.H. Robertson, P. Dyer, T.J. Bowles, R.E. Brown, N. Jarmie, C.J.Maggiore, S.M. Austin, Phys. Rev. C 27 (1983) 11.
\bibitem{vol83} H. Volk, et al., Z. Phys. A 310 (1983) 91.
\bibitem{ale84} T. Alexander, et al., Nucl. Phys. A 427 (1984) 526.
\bibitem{hil88} M. Hilgemeier, et al., Z. Phys. A 329 (1988) 243.
\bibitem{ade98} E.G. Adelberger, et al., Rev. Mod. Phys. 70 (1998) 1265.
\bibitem{fuk01} S. Fukuda, et al., Phys. Rev. Lett. 86 (2001) 5651.
\bibitem{ahm01} Q.R. Ahmad, et al., Phys. Rev. Lett. 87 (2001) 071301.
\bibitem{hos06} J. Hosaka, et al., Phys. Rev. D 73 (2006) 112001.
\bibitem{aha08} B. Aharmim, et al., Phys. Rev. Lett. 101 (2008) 111301.
\bibitem{bel11} G. Bellini, et al., Phys. Rev. Lett. 107 (2011) 141302.
\bibitem{abe11} S. Abe, et al., Phys. Rev. C 84 (2011) 035804.
\bibitem{bah06} J. Bahcall, A.M. Serenelli, S. Basu, Astrophys. J. Suppl. 165 (2006) 400.
\bibitem{wmap} E. Komatsu, et al. [WMAP Collaboration], 2011 Astrophys. J. Supp. 192 (2011) 18.
\bibitem{ioc09} F. Iocco, G. Mangano, G. Miele, O. Pisanti, P.D. Serpico, Phys. Rep. 472 (2009) 1
\bibitem{nar04} B. S. Nara Singh, M. Hass, Y. Nir-El, G. Haquin, Phys. Rev. Lett. 93 (2004) 262503.
\bibitem{bem06} D. Bemmerer, et al., Phys. Rev. Lett. 97 (2006) 122502.
\bibitem{gyu07} Gy. Gy\"urky, et al., Phys. Rev. C 75 (2007) 35805.
\bibitem{con07} F. Confortola, et al., Phys. Rev. C 75 (2007) 065803.
\bibitem{bro07} T.A.D. Brown, et al., Phys. Rev. C 76 (2007) 055801.
\bibitem{cos08} H. Costantini, et al., Nucl. Phys. A 814 (2008) 144.
\bibitem{dil08} A. Di Leva, et al., Phys. Rev. Lett. 102 (2009) 232502.
\bibitem{ade11} E.G. Adelberger, et al., Rev. Mod. Phys. 83 (2011) 195.
\bibitem{kaj84} T. Kajino, A. Arima, Phys. Rev. Lett 52 (1984) 739.
\bibitem{nol01} K.M. Nollett, Phys. Rev. C 63 (2001) 054002.
\bibitem{nef11} T. Neff, Phys. Rev. Lett. 106 (2011) 042502.
\bibitem{tom63} T.A. Tombrello, P.D. Parker, Phys. Rev. 131 (1963) 2582.
\bibitem{kim81} B.T. Kim, T. Izumoto, K. Nagatani, Phys. Rev. C 23 (1981) 33.
\bibitem{buc85} B. Buck, R.A. Baldock, J.A. Rubio, J. Phys. G 11 (1985) L11.
\bibitem{buc88} B. Buck, A.C. Merchant, J. Phys. G 14 (1988) L211.
\bibitem{lan86} K. Langanke, Nucl. Phys. A 457 (1986) 351.
\bibitem{mer86} T. Mertelmeier, H.M. Hofmann, Nucl. Phys. A 459 (1986) 387.
\bibitem{moh93} P. Mohr, et al., Phys. Rev. C 48 (1993) 1420.
\bibitem{dub95} S.B. Dubovichenko, A.V. Dzhazairov-Kakhramanov, Phys. At. Nucl. 58 (1995) 579.
\bibitem{cso00} A. Cs\'ot\'o, K. Langanke, Few-Body Syst. 29 (2000) 121.
\bibitem{des04} P. Descouvemont, A. Adahchour, C. Angulo, A. Coc, E. Vangioni-Flam, At. Data Nucl. Data Tab. 88 (2004) 203.
\bibitem{moh09} P. Mohr, Phys. Rev. C 79 (2009) 065804.
\bibitem{sju13} S.K.L. Sjue, et al., Nucl. Instr. Meth. A 700 (2013) 179.
\bibitem{til02} D.R. Tilley, C.M. Cheves, J.L. Godwin, G.M. Hale, H.M. Hofmann, J.H. Kelley, C.G. Sheu, H.R. Weller, Nucl. Phys. A 708 (2002) 3.
\bibitem{bor12} C. Bordeanu, Gy. Gy\"urky, Z. Elekes, J. Farkas, Zs. F\"ul\"op, Z. Hal\'asz, G.G. Kiss, E. Somorjai, T. Sz\"ucs, Nucl. Instr. Meth A 693 (2012) 220.
\bibitem{cyclotron} http://www.atomki.hu/atomki/Accelerators/Cyclotron/index-en.html
\bibitem{mar06} M. Marta, et al., Nucl. Instr. Meth. A 569 (2006) 727.
\bibitem{gor80} J. G\"orres, et al., Nucl. Instr. and Meth. 177 (1980) 295.
\bibitem{SRIM} http://www.srim.org, version SRIM-2008.04
\bibitem{SRIMerror} http://srim.org/SRIM/SRIM2011.htm
%\bibitem{lem08} A. Lemut, Eur. Phys. J. A 36 (2008) 233.
\bibitem{car12} M. Carmona-Gallardo, et al., Phys. Rev. C 86 (2012) 032801(R).

\end{thebibliography}
\end{document}